# To the identification of universal criterion of the charged particles system stability

Iogann Tolbatov

*Physics and Engineering Department, Kuban State University, Krasnodar, Russia*

(talbot1038@mail.ru)

We consider the problem important for the condensed matter physics, superconductivity physics, and electrodynamics of continuous media - the problem of the matter dielectric permittivity possible values spectrum definition. Two ways of the dielectric permittivity values spectrum identification are analyzed. The proposed technique allows the author to complete the universal criterion of stability identified by D. A. Kirzhnits for a system of charged particles by the criterion of stability for a superconducting system of charged particles.

The problem of dielectric permittivity, its possible values spectrum is very important for the condensed matter physics and electrodynamics of continuous media. This issue is also closely linked with one of the most unexplored areas of physics - the theory of high temperature superconductivity (HTSC). The relationship of dielectric and superconducting properties has been described in the works [1, 2].

The purpose of this article is to show the two ways to identify the range of values of the dielectric constant: the approach used by D. A. Kirzhnits in [3], interpreted by him as a criterion for stability of the system of charged particles, and the approach described in [4], used for the definition of stability criterion of the superconducting system of charged particles.

Both methods include: a) characteristics of the permittivity values range, b) determination of the stability criterion of the charged (superconducting, in the second case) particles, which is to correspond to the dielectric constant.

In the paper [3], D. A. Kirzhnits described $\varepsilon(\vec{q},0)$ as a response function characterizing the causal relationship based on the Kramers-Kronig dispersion [5]. D. A. Kirzhnits justified the stability condition of $\varepsilon(\vec{q},0)$.

In the paper [6], I. V. Tolbatov has entered the condition of the system roughness [6]. On this assumption basis, the search of the two quasi-free electrons forming "Cooper pair" motion equations is performed, these equations have to be functions of coordinates of the two deformation potentials, which determine the local values of the permittivities $\varepsilon_1(x,y)$ and $\varepsilon_2(x,y)$ [4]. Using the rough systems model, he has constructed the charged particles system superconducting state stability condition.

Let us comment on the both authors points of view, compare them, and try by our article to make the necessary refinements to the criteria of the charged particles system stability (in the first case – any system, in the second – the superconducting system) as well as to the method of criteria determination.

1. We consider $\varepsilon(\omega)$ as a function of the response (according to Kirzhnits [3]).

1.1. Constitutive equations describing the environment make the closed system of Maxwell's equations.

Maxwell's equations in the medium have the form:

$$\begin{cases} [\vec{k}\vec{B}] + \omega\vec{E} = -4\pi i \vec{j}; \vec{k}\vec{B} = 0 \\ [\vec{k}\vec{E}] - \omega\vec{B} = 0; \vec{k}\vec{E} = -4\pi i \rho \end{cases},\qquad(1)$$

where $\rho = \rho^i + \rho^e; \vec{j} = \vec{j}^i + \vec{j}^e; \rho^i, \vec{j}^i$ are the induced charge and current densities, appearing under the influence of $\rho^e, \vec{j}^e$ (the external charge and current densities, respectively).

1.2. We enter $\delta\vec{D}; \delta\vec{E}; \delta\vec{B}; \delta\vec{H}$, where $\delta$ is the small variation symbol. Then, the equations (1) transform into the following form:

$$[\vec{k}\vec{H}] + \omega\vec{D} = -4\pi i \vec{j}^e; \vec{k}\vec{B} = 0$$
$$[\vec{k}\vec{E}] - \omega\vec{B} = 0; \vec{k}\vec{D} = -4\pi i \rho^e. \tag{2}$$

Quantities $\vec{D}$ and $\vec{H}$ are uniquely identified with the arbitrariness in the two degrees of freedom.

We introduce the quantities $\alpha$ and $\beta$:

$$\delta\vec{E}_{||} = \alpha\delta\vec{E}^e_{||}; \delta\vec{B} = \beta\delta\vec{H}^e, \tag{3}$$

$$\delta\rho = \alpha\delta\rho^e; \delta\vec{j}_\perp = \beta\delta\vec{j}_\perp^e. \tag{4}$$

Until now, it was logical to consider electric and magnetic fields simultaneously due to greater visibility. However, recalling this article aim, we omit the calculations concerning magnetic field.

We subject the medium to small external influence $(I)$, not necessarily associating it with external source of charges and currents. The result of this external influence is the quantity characterizing the resultant changes in the environment $A$:

$$A = R \times I, \tag{5}$$

where $R$ is the response function describing the reaction of medium on the external impact without regard to its amplitude.

Let us determine whether the quantities $\alpha$ and its inverse are the response functions. A comparison of (3) and (5) implies that

$$\alpha = R, \tag{6}$$

where $\delta\vec{E}^e$ is the external field; and the quantity $A$ is the full field $\delta\vec{E}$. Its inverse quantity also is the response function:

$$\frac{1}{\alpha} = R. \tag{7}$$

1.3. Identifying the dispersion relations, we find that the response function satisfies the causality. In the case of the longitudinal electric field, we consider the non-relativistic causality condition: the result of the impact $A$ is zero at all times prior to the influence $I$. We apply the Cauchy formula to the function $R(\omega, \vec{k}) - R(\Omega, \vec{k})$ (if $\Omega \to 0$), and the Kramers-Kronig relation transforms into:

$$R(\omega, \vec{k}) = R(\Omega, \vec{k}) + \frac{1}{\pi} \int_0^\infty \frac{\operatorname{Im} R(\omega', \vec{k})}{\omega'^2 - \omega^2 - i\delta} d\omega'^2. \tag{8}$$

Since $R \sim |\vec{k}|$, then $\vec{k} - \vec{u}\omega \to (k^2 + \omega^2)^{1/2}$, in this case, it follows from (8) that

$$R(0, \vec{k}) = R(\Omega, \vec{k}) + \frac{1}{\pi} \int_0^\infty \operatorname{Im} R(\omega', \vec{k}) \frac{d\omega'^2}{\omega'^2}. \tag{9}$$

1.4. It follows from [7] that the monochromatic wave energy dissipation per unit time has the form:

$$Q = \frac{\omega}{8\pi}((\operatorname{Im}\varepsilon_{||})E_{||}^2)^2 + \operatorname{Im}\varepsilon_\perp (E_\perp)^2) \geq 0$$

$$\Rightarrow \operatorname{Im}\varepsilon_{||} \geq 0; \operatorname{Im}\varepsilon_\perp \geq 0.$$

Hence, $\operatorname{Im}\alpha \leq 0$. (10)

Asymptotics of $\alpha$ at $\omega \to \infty$ follows from the fact that the environment does not have time to respond to the impact of a frequency greater than its characteristic frequency.

$$\omega \to \infty \Rightarrow \alpha(\Omega, \vec{k}) = 1 \tag{11}$$

1.5. Acceptable values for the permittivity are:

$$(6,10,11,9) \to \alpha(0, \vec{k}) = \frac{1}{\varepsilon(0, \vec{k})} \leq 1 (\vec{k} = \forall). \tag{12}$$

Substituting (7) instead of (4), we obtain:

$$\frac{1}{\alpha(0, \vec{k})} = \varepsilon(0, \vec{k}) \geq 1 (\vec{k} = 0) \tag{13}$$

1.6. Now we show the condition, under which the system of charged particles is stable. If $I$ is the external influence, $\Phi$ is the physical quantity fluctuation, and $F(I)$ is the free energy variation, then

$$\delta F(I) = \Phi \delta I.$$

Using the Legendre transformation and the stability criterion of Le Chatelier and Brown, we obtain [3]:

$$\frac{\delta(\Phi(I) - \Phi_I)}{\delta I} \leq 0.$$

Let $\delta F = E \dfrac{\delta D}{4\pi}; I = \dfrac{\vec{D}}{4\pi}; \Phi = \vec{E}; \Phi_I = \vec{D}$, then

$$\frac{\delta(\vec{E} - \vec{D})}{\delta \vec{D}} = \alpha(0, \vec{k}) - 1 \leq 0 \text{ coincides with (12).}$$

Thus, Kirzhnits condition for stability of charged particles takes the form:

$$\frac{1}{\varepsilon(\vec{q}, 0)} < 1.$$

2.1. Using a different approach described in [4], we consider the system of the charged particles, on which external forces do not influence, that is: the particles either rest or move without acceleration (there is no second derivative on time):

$$\frac{dx}{dt} = P(x, y); \quad \frac{dy}{dt} = Q(x, y); \tag{14}$$

where $x = x(t)$, $y = y(t)$ – particles trajectories, $P(x, y)$, $Q(x, y)$ – analytical functions.

Dependence of particle trajectories from time $t$ indicates the presence of the so-called "cycle without contact" because in the absence of external (and dissipative) forces, the total energy $E$ is constant, and the total work of force acting on a particle is: $\oint_L A dl = 0$, and the trajectory $L$ is closed. In this case, there exists a domain $G$ bounded by simple closed curve $g$ with a continuously rotating tangent. Outside of $G$, the system has no "cycle without contact".

Now we consider alongside with the system (14) the changed system:

$$\frac{dx}{dt} = P(x,y) + p(x,y); \qquad \frac{dy}{dt} = Q(x,y) + q(x,y); \qquad (15)$$

where *p(x,y)* and *q(x,y)* are the analytical functions, for which the following conditions are true:

$$\forall \delta > 0; \quad \forall \varepsilon > 0$$

$$\exists \forall p(x,y): \ |p(x,y)| < \varepsilon; \ |p'_x(x,y)| < \varepsilon; \ |p'_y(x,y)| < \varepsilon;$$

$$\exists \forall q(x,y): \ |q(x,y)| < \varepsilon; \ |q'_x(x,y)| < \varepsilon; \ |q'_y(x,y)| < \varepsilon.$$

There is a mutually unique and mutually continuous transformation of *K* of *G* domain into itself in which: 1) relevant to each other points are at a distance of less than $\delta$, 2) the points lying on the same trajectory of system (14) correspond to the points lying on the same trajectory of system (15), and vice versa. Thus, in *G*, the system (14) is a rough system.

If the system (14) is rough in *G*, then in *G*, the system (14) can only have such an equilibrium position, for which the real parts of the roots of the characteristic equation are different from zero.

Consequently, the system (14) in *G* can not have equilibrium $x = x_0$ and $y = y_0$, for which the following is true:

a) $\Delta = 0$, (16)

b) if $\Delta > 0$ $\sigma = -(P'_x(x_0, y_0) + Q'_y(x_0, y_0)) = 0$, (17)

where $\Delta = \begin{vmatrix} P'_x(x_0,y_0) & P'_y(x_0,y_0) \\ Q'_x(x_0,y_0) & Q'_y(x_0,y_0) \end{vmatrix}.$ (18)

We find the stability criterion for a system of electrons. The potential of electron-electron interaction has the property of asymptotic freedom, i. e. the total effect of the Coulomb potential and the potential of deformation of the crystal lattice are such that the electrons of the Cooper pairs move quasifree. Therefore, they should be regarded as free electrons, then the equation of electron motion in electric field of an external monochromatic electromagnetic wave has the form [5]:

$$m\ddot{x} = eE_0 e^{-i\omega t}. \tag{19}$$

From the equation (19), we obtain the law of the frequency dispersion of the electron system permittivity in a standard way:

$$\varepsilon(\omega) = 1 - \frac{\omega_p^2}{\omega^2}, \tag{20}$$

where $\omega_p$ is the frequency of plasma oscillations.

Expressing from (20) the oscillation frequency $\omega = \frac{\omega_p}{\sqrt{1-\varepsilon}}$ through the dielectric constant and substituting the latter in (19), we find the equations of motion of two quasifree electrons forming a Cooper pair as the functions of the coordinates of the two deformation potentials, which determine the local values of the permittivities $\varepsilon_1(x,y)$ and $\varepsilon_2(x,y)$ corresponding to the electron localization coordinates $x$ and $y$.

$$\frac{dx}{dt} = \frac{ieE_0}{m\omega_p}\sqrt{1-\varepsilon_1(x,y)}\exp\left(-\frac{i\omega_p t}{\sqrt{1-\varepsilon_1(x,y)}}\right), \tag{21}$$

$$\frac{dy}{dt} = \frac{ieE_0}{m\omega_p}\sqrt{1-\varepsilon_2(x,y)}\exp\left(-\frac{i\omega_p t}{\sqrt{1-\varepsilon_2(x,y)}}\right).$$

Calculating the discriminant $\Delta = \begin{vmatrix} P'_x(x_0,y_0) & P'_y(x_0,y_0) \\ Q'_x(x_0,y_0) & Q'_y(x_0,y_0) \end{vmatrix}$, we find the stability criterion from relations (14) and (21).

Thus, the criterion of stability of the charged particles system has the form:

$$\begin{bmatrix} a)\,\varepsilon'_{1x}\varepsilon'_{2y} > \varepsilon'_{1y}\varepsilon'_{2x}; \dfrac{\varepsilon'_{1x}}{\varepsilon'_{2y}} > 0; \varepsilon_{1,2} < 1 \\ \text{б})\,\varepsilon'_{1x}\varepsilon'_{2y} < \varepsilon'_{1y}\varepsilon'_{2x}; \varepsilon_{1,2} < 1 \end{bmatrix}$$

So, we have shown two approaches to the problem of defining the charged particles system stability criterion: the approach used by Kirzhnits [3], and the approach studied in [4].

Criterion $\frac{1}{\varepsilon(\vec{q},0)} < 1$ (K)

is the condition for stability of the system of charged particles.

Criterion $\begin{bmatrix} a) \varepsilon'_{1x}\varepsilon'_{2y} > \varepsilon'_{1y}\varepsilon'_{2x}; \frac{\varepsilon'_{1x}}{\varepsilon'_{2y}} > 0; \varepsilon_{1,2} < 1 \\ 6) \varepsilon'_{1x}\varepsilon'_{2y} < \varepsilon'_{1y}\varepsilon'_{2x}; \varepsilon_{1,2} < 1 \end{bmatrix}$ (T)

is the condition of stability of the superconductivity state in the matter. It can be interpreted as a condition for the stability of "Cooper pair" - the superconducting system of charged particles. We see that the criterion (T) is a special completion to the criterion (K), since inequality $\frac{1}{\varepsilon(\vec{q},0)} < 1$ splits into two inequalities $\begin{bmatrix} \varepsilon(\vec{q},0) > 1 \\ \varepsilon(\vec{q},0) < 0 \end{bmatrix}$, where the inequality $\varepsilon(\vec{q},0) < 0$ corresponds to the superconducting state of matter (that follows from a comparison of (K) and (T)), and it is a necessary but not sufficient condition. Thus, we see the incompleteness of each of the criteria and, due to our analysis, we obtain the refined criteria:

1) criterion (K) $\frac{1}{\varepsilon(\vec{q},0)} < 1$ is completed by the criterion (T) in the superconducting state case,

2) criterion (T) is converted to a form $\begin{bmatrix} a) \varepsilon'_{1x}\varepsilon'_{2y} > \varepsilon'_{1y}\varepsilon'_{2x}; \frac{\varepsilon'_{1x}}{\varepsilon'_{2y}} > 0; \varepsilon_{1,2} < 0 \\ 6) \varepsilon'_{1x}\varepsilon'_{2y} < \varepsilon'_{1y}\varepsilon'_{2x}; \varepsilon_{1,2} < 0 \end{bmatrix}$, where the inequality $\varepsilon_{1,2} < 0$ is postulated in (K).

In our study, we have shown two approaches to the stability criterion definition: in one case - the universal criterion for systems of charged particles, in other case – the criterion for the special (superconducting) situation. The criteria were refined and completed. It can be also concluded that the criterion

(T) is a special case of (K), but very important as it encompasses the wide range of superconducting materials.

## REFERENCES

1. Bardeen J., Cooper L.N., Schrieffer J.R. // Phys. Rev. 108 1175 (1957).
2. Cohen M.L., Anderson P.W. // Superconductivity in d- and f- Band Metals (AIP Conf. Proc., Ed D H Douglass) (New York: AIP, 1972). P. 17.
3. Kirzhnits D. A. / / UFN. 119 367 (1976).
4. Criterion of stability of the superconducting state. / Tolbatov I. V. / Journal Izvestiya VUZov. Pish. Tech.; Krasnodar, 2007; 11 p., bibl. 9.; Rus.; Published in VINITI RAN 07.11.2007, N 1039 B2007.
5. Bredow M. M., Rumyantsev V. V., Toptygin I. N. - Classical Electrodynamics. - SPb.: Lan', 2003.
6. Pontryagin L. S., Andronov A. A. // Dokl. AN SSSR. - 1937. - Vol. 14, N 5. - p. 247-250.
7. Agranovich V. M., Ginzburg V.L. Crystal Optics with Spatial Dispersion and Excitons. - M.: Nauka, 1979.